\documentclass{article}
\usepackage{arxiv}
\usepackage[utf8]{inputenc} 
\usepackage[T1]{fontenc}    
\usepackage{hyperref}       
\usepackage{url}            
\usepackage{booktabs}       
\usepackage{amsfonts}       
\usepackage{nicefrac}       
\usepackage{microtype}      
\usepackage{graphicx}
\usepackage{doi}
\usepackage{bm}
\usepackage{array}
\usepackage{eqparbox}
\usepackage{commath}
\usepackage{nicefrac}    
\usepackage{microtype} 
\usepackage[section]{placeins}
\usepackage{hhline}
\usepackage{cite}
\usepackage{subfigure}
\usepackage{multirow}
\usepackage{array}
\usepackage{mdwmath}
\usepackage{mdwtab}
\usepackage{amssymb,amsmath,amsthm,mathtools}
\usepackage{enumerate}
\usepackage{setspace}
\usepackage{comment}
\usepackage{cleveref}
\usepackage{float}
\usepackage{makeidx}
\usepackage{algorithm2e}
\usepackage{xcolor}

\title{A Probabilistic Approach to Adaptive Protection in the Smart Grid}

\date{} 					

\author{{Amr S.~Mohamed} \hspace{7mm} {Deepa~Kundur}\\
	Department of Electrical Engineering\\
	University of Toronto\\
	Toronto, ON M5S 3G4, Canada \\
	\texttt{amr.mohamed@mail.utoronto.ca} \\
 \texttt{dkundur@ece.utoronto.ca} \\
        \And
	Mohsen Khalaf \\
         Department of Electrical Engineering\\
	University of Toronto\\
 Toronto, ON M5S 3G4, Canada \\
	and Assiut University\\
	Assiut, Egypt\\
	\texttt{m.khalaf@utoronto.ca}\\
        \texttt{mohsen.b.khalaf@aun.edu.eg} \\
}

\begin{document}
\maketitle

\begin{abstract}
Smart grids are critical cyber-physical systems that are vital to our energy future. 
Smart grids' fault resilience is dependent on the use of advanced protection systems that can reliably adapt to changing grid conditions. 
The vast amount of operational data generated and collected in smart grids can be used to develop these protection systems.
However, given the safety-criticality of protection, the algorithms used to analyze this data must be stable, transparent, and easily interpretable to ensure the reliability of the protection decisions. Additionally, the protection decisions must be fast, selective, simple, and reliable. To address these challenges, this paper proposes a data-driven protection strategy, based on Gaussian Discriminant Analysis, for fault detection and isolation. This strategy minimizes the communication requirements for time-inverse relays, facilitates their coordination, and optimizes their settings.
The interpretability of the protection decisions is a key focus of this paper. The method is demonstrated by showing how it can protect the medium-voltage CIGRE network as it transitions between islanded and grid-connected modes, and radial and mesh topologies.
\end{abstract}

\keywords{Adaptive Protection, Microgrid, Gaussian Discriminant Analysis, Machine Learning, Inverse-time Protection}

\section{Introduction} \label{sec:intro}

The smart grid is a critical cyber-physical system that plays a central role in modernizing the energy landscape. It stands to enhance energy security, facilitate the integration of renewable energy sources, and enable significant improvements in energy efficiency.
Enabling the smart grid is dependent on the development of reliable protection schemes to ensure its resilience against faults. 
Smart grid protection is a highly challenging task, particularly at the distribution level in microgrids, due to factors such as bidirectional power flow, flexible topology, and the increasing utilization of renewable energy sources~\cite{brearley2017review, lin2019adaptive}. As microgrids transition between different topology configurations and between grid-connected mode with a high short-circuit current capacity and islanded mode with limited and variable capacity, it is vital for the protection schemes to adapt to changing grid conditions~\cite{hooshyar2017microgrid, barra2020survey}.

Adaptive protection is regarded as the most suitable approach to protecting microgrids \cite{brearley2017review, senarathna2019review}. It involves changes in the fault response of protection relays based on the system operating state and conditions  \cite{senarathna2019review, khalid2021existing}. 
This is commonly achieved by extending the functionality of existing traditional protection devices, such as inverse-time over-current relays, which constitutes a cost-effectiveness benefit in the absence of specialized commercial microgrid relays \cite{hooshyar2017microgrid} and the desire to easily integrate protection schemes into existing devices and substations with minimal capital expenditure \cite{abdulhadi2011adaptive}. 

Supported by years of operation in traditional power systems, over-current relays offer an economical and effective solution to protect the power system against faults. 
With the rise of microgrids, there is a growing interest to extend the capabilities of over-current relays to enable adaptive protection \cite{bhattarai_adaptive_2015, wong2021selectivity, yousaf2021control}. This requires the development of methods to update over-current relay settings in response to changes in the grid, while maintaining both selectivity and sensitivity \cite{habib2017adaptive}. Selectivity refers to the relay's ability to precisely locate and classify faults to isolate the smallest possible faulty section of the power system, while sensitivity refers to the ability to promptly and accurately detect faults to quickly clear them.

Recent studies have explored the use of data-driven techniques, including machine learning algorithms, to leverage the growing amount of data generated in power systems in satisfying these protection requirements. 
Data-driven techniques have been applied to detecting faults \cite{lin2016adaptive, lin2019adaptive}, identifying the location, type, and phase of faults \cite{james2017intelligent, abdelgayed2017new}, and adjusting relay settings \cite{tang2018data}. 
Based on the findings of these studies, it appears that data-driven techniques can offer several benefits.
Firstly, data-driven methods can enhance the accuracy of protection decisions, improving fault detection and classification and reducing unnecessary tripping. 
For example, Mishra \textit{et al.} \cite{mishra2015combined} conducted a study where they compared the accuracy of fault detection in protection schemes based on decision trees and random forests with traditional over-current relays. Their study showed over 40\% improvement in fault detection accuracy when leveraging the machine learning algorithms.
Secondly, data-driven techniques can offer a scalable solution to adapting relay settings, significantly reducing the need to manually identify system topology configurations and compute relay settings. 
For example, in centralized communication-based implementations, data-driven techniques have the potential to replace traditional relays, monitoring the entire power system and directly controlling circuit breakers \cite{senarathna2019review}.
We survey some of the literature applying data-driven techniques to adaptive protection in Section \ref{sec:lit_review}.


Data-driven techniques, however, come with their own set of limitations and potential threats to protection reliability. 
Black-box algorithms, such as neural networks and random forests, while often highly accurate, cannot be easily verified or validated, making them a significant risk for adoption in safety-critical applications such as protection.
Some other algorithms such as decision trees, while more interpretable, may suffer from instability: inconsistent decisions resulting from small changes in input data. This instability can also make it difficult to validate their decisions, thus undermining their feasibility for protection.
To successfully integrate data-driven techniques into adaptive protection, it is crucial to ensure that the interpretability of the protection decisions is preserved. This will enable protection engineers to verify the decisions while simultaneously reducing potential threats to reliability.

Additionally, adaptive protection schemes, which rely on real-time communication infrastructure such as peer-to-peer communication between relays or communication between a central protection management center and relays, may face potential threats to reliability due to network failure, latency, and cyberattacks \cite{gutierrez2020review, habib2017adaptive}. Although increased communication can lead to more accurate and informed protection decisions, the associated risks and threats must be carefully considered.

To acquire benefits of data-driven techniques and communication-based adaptive protection while reducing threats to reliability, we propose a novel data-driven adaptive protection scheme that emphasizes the interpretability of protection decisions for validation and minimizes reliance on communication.
Our proposed approach is based on Gaussian discriminant analysis and is designed to: (1) identify the system topology configurations that require a change in the protection relays' fault response, (2) specify necessary communication with relays, (3) locate faults with respect to relays for relay coordination, and (4) optimize relay settings for fast fault clearing. This approach is centralized with a central microgrid protection management center communicating with relays following infrequent major grid changes.
To demonstrate the effectiveness of our proposed strategy, we apply it to inverse-time protection relays.

\textbf{Contributions.} 
\begin{itemize}
    \item Our work presents a first-of-its-kind application of Gaussian discriminant analysis (GDA) to adaptive protection. Through our research, we discuss and demonstrate several benefits of GDA for adaptive protection, including improved data- and computational-efficiency, interpretability for ease of validation, adaptability to changing grid conditions, timely response, and reduced reliance on communication for protection.
    To showcase the interpretability of our proposed strategy, we put emphasis on visualizing its protection decision-making process in this paper. 

    \item We extend the inverse-time protection formula to improve fault distinction by incorporating voltage drops associated with faults. 

    \item We formulate an optimization problem that enables the coordination and optimization of relays in a changing system environment.
\end{itemize}

We validate the strategy on the CIGRE medium voltage benchmark system. 
The benchmark system permits many configurations to help validate the proposed adaptive protection method, including radial and mesh topology, and grid-connected and islanded modes. 

\textbf{Organization. }The rest of the paper is structured as follows: 
We survey related work in Section \ref{sec:lit_review}.
We present background material in Section \ref{sec:background}.
We present the problem statement discussing the goals of data-driven adaptive protection in Section \ref{sec:problem_formulation}.
We develop the protection strategy in Section \ref{sec:method}.
Next, we illustrate and discuss the results of the protection strategy using the benchmark system in Section \ref{sec:results}. The conclusion is in Section \ref{sec:conclusion}.

\section{Related Work} \label{sec:lit_review}

The literature reviews in \cite{gutierrez2020review, barra2020survey, khalid2021existing} provide different perspectives on adaptive protection. We will summarize the contents of these reviews that are relevant to our work. Our work proposes an intelligent centralized protection scheme with pre-set relays, which involves a central microgrid protection management center issuing instructions to the relays to switch their settings only after significant topology changes. 
Next, we will survey some of the literature that applies data-driven techniques to adaptive protection to motivate our work.

In their analysis of literature on adaptive protection in the context of communication, Gutierrez-Rojas \textit{et al.} \cite{gutierrez2020review} categorize their examination according to communication technology (wireless or wired) and control approach (centralized or decentralized). Most of the adaptive protection research proposes centralized control approaches. In centralized control, a central control center is established to monitor the power system in real-time and send (either wired or wireless) instructions to relays to switch between relay settings or directly to circuit breakers to trip. For example, Oudalov \textit{et al.} \cite{oudalov2009adaptive} implemented a centralized control approach in which wired communication is established between the control center and advanced circuit breakers with integrated over-current relay units. In real-time, data is collected from the circuit breakers and relay settings are modified. Alternatively, communication can be limited to notifying relays of infrequent major grid topology changes, such as network reconfiguration, as in \cite{bhattarai_adaptive_2015}, with local measurements used to detect other changes, such as distributed energy resource status.

Barra \textit{et al.} \cite{barra2020survey} organize their review by the methods that rely on computational intelligence, such as machine learning, fuzzy logic, and optimization, and methods that do not.

Khalid \textit{et al.} \cite{khalid2021existing} categorize their review based on papers that propose methods for fault detection, relay setting adjustment to changes in network configuration, or relay optimization for coordination, and papers with methods that require and do not require pre-setting relays.

Recent studies have explored the use of data-driven techniques to make use of the growing streams of data in power systems for more accurate and reliable adaptive protection. 
In these studies, machine learning algorithms are used to identify (or detect) the presence of a fault, classify the type of fault (phase-to-ground, phase-to-phase, etc.), and locate the fault in order to isolate it. 
Typically, the data collection, pre-processing, and mining, as well as model training are performed offline. 
For example, Lin \textit{et al.} in \cite{lin2016adaptive} trained two artificial neural networks (ANN) models to classify and locate faults using current measurements. 
The fault localization ANN was replaced with a support vector machine (SVM) in \cite{lin2019adaptive}. 
In \cite{lin2019adaptive}, the fault localization accuracy of the SVM model was consistently over 99\% (when testing the method using the IEEE 9 bus system). However, the classification accuracy of the ANN dropped to 77.5\% and 80.5\% on some lines where fault classification was challenging.
This highlights the challenge of accurately classifying faults in microgrids. 

Similarly, Zayandehroodi \textit{et al.} in \cite{zayandehroodi2012novel} used combinations of radial-basis function ANNs to locate faults using current measurements. 
Their results show that the first ANN was able to locate the fault within 0.01 km error. The second ANN predicts the faulty line based on the fault location.
They propose a backtracking algorithm to coordinate between relays.

Other studies like James \textit{et al.} \cite{james2017intelligent} and Tang \textit{et al.} \cite{tang2018data} trained algorithms on statistical features extracted from wavelet transforms of current measurements. 
James \textit{et al.} used deep recurrent ANNs with to classify and locate faults. Their method achieved a fault detection, and fault-type and -phase classification accuracies of over 97\%, and over 94\% accuracy in fault localization.
Tang \textit{et al.} used decision trees (DT) local to each relay to classify faults. Instead of using communication with a central controller, they proposed a peer-to-peer communication system between relay intelligent electronic devices (IED).
A novelty in Tang \textit{et al.} was the use of an ANN to adjust relay settings.

Although neural networks can be highly accurate and valuable for addressing the complexities of adaptive protection, they pose significant risks for adoption due to their black-box nature and difficulties in verification and validation. 
Alternatively, the authors in \cite{abdelgayed2017new}
and \cite{mishra2015combined} applied interpretable machine learning algorithms to adaptive protection.
Abdelgayed \textit{et al.} \cite{abdelgayed2017new} applied a variety of algorithms to statistical features extracted from wavelet transforms of current measurements to classify faults type. The accuracy of the algorithms were as follows: k-nearest neighbors (kNN) with 95.63\%, DT with 90.4\%, SVM with 93.3\%, and na\"ive Bayes with 94.24\%.
Mishra \textit{et al.} used DTs to detect and classify faults with 97\% and 85\% accuracy, respectively.

Although their decisions are interpretable, DTs have a drawback of instability, meaning that small changes in the training data can result in large changes in the structure of the decision tree, leading to vastly different hypotheses and undermining their interpretability \cite{kenneth2007decision}.
The kNN algorithm is slow in test time as it involves calculating the distance of the test instance to all training data. This means that as the size and dimensionality of the training data increases, the computation time for making a prediction also increases. This can make kNN impractical for real-time applications with large datasets that require timeliness. 
Na\"ive Bayes and GDA are both probabilistic data-driven techniques that are similar. GDA can learn more complex features and potentially achieve higher accuracy.
We note that probabilistic data-driven methods are generally stable, intuitive, interpretable, and computationally- and data-efficient \cite{ghahramani2015probabilistic, ghahramani2017pres}.
SVM and GDA are comparable; we discuss why we apply GDA over SVM in Section \ref{sec:results}.

\section{Background} \label{sec:background}

{We apply our adaptive protection method to the optimization and coordination of inverse-time protection relays. 
We begin this section by describing the operation of time-inverse relays. 
Next, we introduce GDA, which is the probabilistic data-driven approach that we use to distill fault data into protection decisions. 
Lastly, we introduce Conditional Value-at-risk, which we use in parametrizing the fault data for relay optimization.}

\subsection{Inverse-time Protection} \label{sec:adaptiveprotection}

Over-current relays apply the following inverse-time equation \cite{benmouyal1999ieee} to compute their operation time: the delay before the relay trips. 
\begin{equation} \label{eq:toc}
    t_{op}(I_M; \text{TMS}, \eta, \lambda, L, I_S) = \text{TMS} \left( \frac{\lambda}{\left( \frac{I_M}{I_S} \right)^\eta - 1} + L \right)
\end{equation}
where $t_{op} > 0$ is the relay operation time, $I_M$ is the relay current measurement, and TMS and $I_S$ are the time-multiplier and pick-up current settings of the relay, respectively. 
Directional over-current relays measure current in one direction making them useful for protecting mesh and multi-source grids.

Equation (\ref{eq:toc}) defines a curve that is shaped by characteristic constants  $\lambda$, $\eta$, and $L$.
Adaptive protection involves shifting the curve to cope with changing grid conditions \cite{che2014adaptive}.

Inverse-time over-current relays work on the principle that fault currents decrease with increasing distance to the fault location. 
As such, shorter operating times are computed per (\ref{eq:toc}) for nearer faults. This facilitates relay coordination where relays closer to the fault location trip earlier to isolate the smallest possible section of the power system.

Anticipating the challenge of coordinating relays only based on fault currents in distribution systems, such as microgrids, due to limited fault currents, we extend \eqref{eq:toc} to consider fault voltage.
\begin{align} \label{eq:tocV}
    t_{op}&(I_M, V_M; \text{TMS}, \eta_i, \eta_v, \lambda_i, \lambda_v, L_i, L_v, I_S, V_S) = \\ 
    &\text{TMS} \left( \frac{\lambda_i}{\left( \frac{I_M}{I_S} \right)^{\eta_i} - 1} + L_i \right) \left( \frac{\lambda_v}{\left( \frac{V_S}{V_M} \right)^{\eta_v} - 1} + L_v \right) \nonumber
\end{align}
where $V_M$ is the relay voltage measurement and $V_S$ is the pick-up voltage settings of the relay. Setting $\lambda_v=0$ and $L_v=1$ in \eqref{eq:tocV} yields \eqref{eq:toc}.

In \eqref{eq:tocV}, the trip time is inversely proportional to the current magnitude and inversely proportional to the voltage magnitude. This means that the protective device will trip faster for larger currents and voltage drops and slower for smaller currents and voltage drops.

\subsection{Gaussian Discriminant Analysis}

GDA is a probabilistic data classification algorithm that fits a Gaussian distribution to each class of data. 
Let $y \in \{1,2,...,n\}$ represent the classes, and $n$ be the number of classes. 
The probability of sampling an observation (or measurement) $\bm{x}$ given class $y = k$ is
\begin{equation} \label{eq:1}
    P(\bm{x}|y=k) = \mathcal{N} (\bm{x}; \bm{\mu}_k, \bm{\Sigma}_k)
\end{equation}
where $\bm{\mu}_k$ and $\bm{\Sigma}_k$ are the mean vector and covariance matrix of the Gaussian distribution fitted to class $k$, respectively.
A GDA classifier {infers} the observation's class via Bayes' theorem: 
\begin{equation} \label{eq:2}
    P(y=k|\bm{x}) = \frac{P(\bm{x}|y=k) P(y=k)}{\sum_{i=1}^{n} P(\bm{x}|y=i) P(y=i)}
\end{equation}


GDA can also be used for supervised dimensionality reduction to project the data to a reduced subspace, which maximizes the separation between classes. 
Following an assumption that all classes share the same covariance matrix of $\bm{\Sigma}_y = \bm{\Sigma}$ for $y = \{1,2,...,n\}$, we can choose this reduced subspace to be linear, which aids in interpreting the results of GDA. The resulting algorithm is termed Linear Discriminant Analysis (LDA). Taking the log-posterior of (\ref{eq:2}) gives the following equation
\begin{equation}
    \log P(y=k|\bm{x}) = (\bm{\Sigma}^{-1} \bm{\mu}_k) \bm{x} + C
\end{equation}
where the coefficients in $\bm{\Sigma}^{-1} \bm{\mu}_k$ determine the importance of the observations' features in separating the classes ($C$ includes other unimportant terms).
Below a certain threshold, unimportant features can be neglected to reduce the dimensionality of the problem.

\subsection{Conditional Value-at-risk}

\begin{figure}
    \centering
    \includegraphics[width=0.5\textwidth, trim={0cm, 6.5cm, 0cm, 6.5cm}, clip]{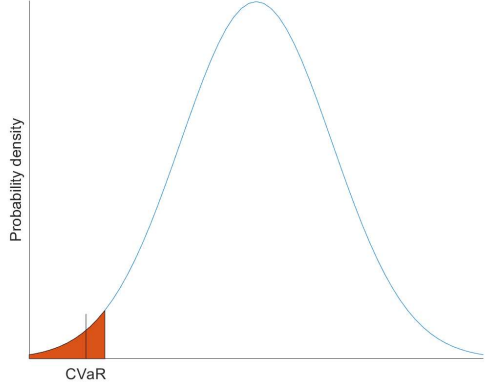}
    \caption{Illustrating the CVaR of a Gaussian distribution. The red area under the curve represents the $100\alpha \%$ tail of the Gaussian distribution. The CVaR$_{\alpha}$ value is the mean of this area and is represented by the vertical line.}
    \label{fig:cvarIll}
\end{figure}

Conditional Value-at-risk (CVaR) \cite{rockafellar2002conditional} is a risk measure of the expected value of a random variable in the lowest $100\alpha \%$ of cases. 
Let $X$ be the random variable and $q_\alpha$ be the $100\alpha$ percentile value of the distribution, then the CVaR with a confidence interval of $100\alpha \%$ is
\begin{equation}
    \text{CVaR}_{\alpha} = E(X | X \leq q_\alpha)
\end{equation}

For a Gaussian distribution, CVaR has the following closed-form:
\begin{equation}
    \text{CVaR}_{\alpha} = \mu - \sigma \frac{\phi(\Phi^{-1}(\alpha))}{\alpha}
\end{equation}
where $\mu$ and $\sigma$ are the distribution's mean and standard deviation, respectively; $\phi$ is the density function; and $\Phi^{-1}$ is the standard normal quantile. CVaR$_{1}$ is equivalent to the mean of the random variable

Fig. \ref{fig:cvarIll} demonstrates the area representing the $100\alpha \%$ tail of the normal distribution in red. CVaR$_{\alpha}$ is the mean of this area.

\section{{Problem Statement}} \label{sec:problem_formulation}

{
Our data-driven adaptive protection method distills data into protection decisions that address the following goals: (1) fault identification and classification, (2) relay optimization and coordination, and (3) communication specification. This section discusses these goals.
}

{
When a power system experiences a short-circuit fault, the sudden reduction in impedance due to the introduction of the short-circuit causes a current rush. 
Over-current relays must measure the increased current and isolate the fault to prevent significant consequences, including power system instability or equipment damage. 
Primary relays, closest to the fault, must trip earliest to isolate the smallest possible section of the power system, with backup relays coordinated to trip in case the primary fails. 
In practice, primary relays should operate following a short delay to allow temporary faults to clear, and backup relays should operate following a preset minimum coordination time (MCT) delay to coordinate with the downstream relay.
}

{
Hence, in relation to a relay, faults can be classified according to their distance from the relay.
A fault is primary (to a relay) if the relay is the closest device to the fault location.
The next (closest) relay(s) are backup to the primary relay. 
A fault can have multiple primary relays and a relay can have multiple backups in situations of equidistant relays from the fault.
In this paper, we use the terms secondary and tertiary faults to classify faults that are progressively further from a relay, and \textit{`other'} to collectively classify tertiary and further faults. {Fault identification and classification} involve enabling relays to accurately identify the existence and class of a fault from local current and voltage measurements.  
Identifying the class of the fault helps relays locate the fault, which enables their coordination.
{Relay optimization and coordination} involve computing relay settings to ensure rapid, correct, selective, and coordinated protection. 
}

{
Variations in power system operating conditions can significantly change the currents measured by relays, which can cause relay misoperation. 
Relays must adjust their operation in response to major grid changes, including the transition between grid-connected and islanded modes of operation, grid topology, and generator availability, which can significantly affect fault current. 
This involves communicating (typically per the IEC61850 \cite{iec61850:2021_2021} protocol) these grid changes to the relays and directing them to switch their relay settings.   
It is desirable to minimize the reliance on communication to reduce the infrastructure capital costs and system complexity and prevent risks associated with communication delays and cyberattacks. {Relay communication specification} involves specifying the contents of the information that will be communicated to the relays, and the communication channels required to enable reliable protection. 
}

{
Finally, a comprehensive protection strategy must consider immeasurable, uncertain factors, including fault impedance, renewable energy intermittency, and fault distance from relays, which also influence the fault current. 
}

The next section will introduce a GDA-based method that will address these goals.

\section{Proposed Adaptive Protection Method} \label{sec:method}

\textbf{Terminology.} We introduce the following {terminology} before presenting the method: 
We use $P(X|Y;Z)$ to denote the probability of a random variable $X$ given $Y$. 
The distribution parameters are $Z$. 
We denote directional relays with R$A$-$B$, where $A$ is the bus it is located at and $B$ is the bus it is directed to. 
Each relay stores multiple relay setting groups. A setting group specifies the TMS, characteristic constants, and pick-up settings used by the relay to compute its operation time. 
Each group is for a distinct relay mode of operation, which is dependent on the state of the system.
The superscript on variable $X^{(r)}$ associates it with a specific relay $r$ and the uppercase superscript on variable $X^{(Z)}$ additionally associates it with a specific relay mode $\bm{Z}$. The subscript $\psi$ denotes the fault class (e.g., primary fault).

\begin{figure*}
    \centering
    \includegraphics[width=\textwidth]{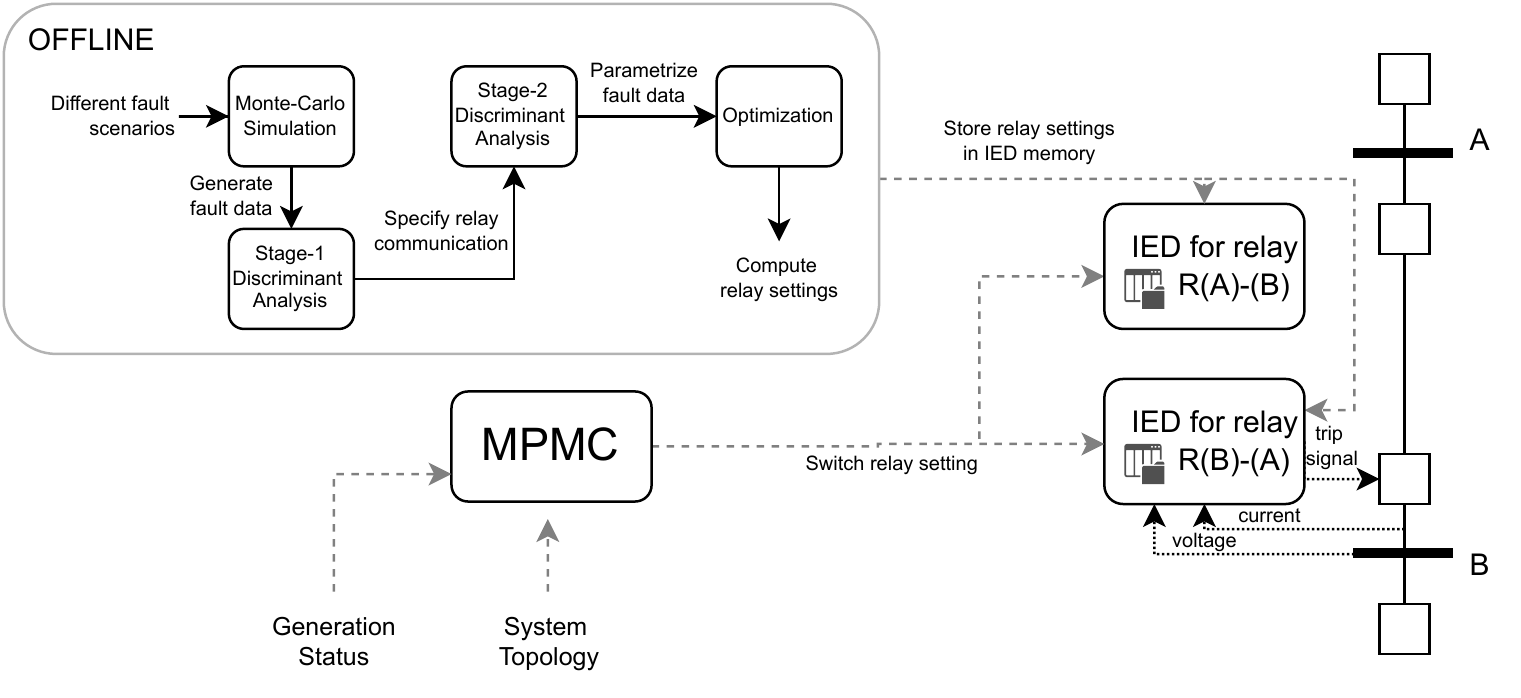}
    \caption{Flowchart of the proposed method.}
    \label{fig:APscheme}
\end{figure*}

The proposed method consists of the steps shown in Fig. \ref{fig:APscheme}, which aim to address the goals outlined in the Problem Formulation, Section \ref{sec:problem_formulation}. 
The relay setting groups are computed via offline steps prior to deployment.
The offline steps consist of running Monte Carlo simulations to generate fault data representing the various fault scenarios that a protection relay may encounter following deployment.
The fault data consists of local current and voltage measurements that are locally accessible to the relay and discrete system states, including switch and generation states, that can be communicated to the relay from the central Microgrid Protection Management Center (MPMC) when needed.
Next, we apply the dimensionality reduction utility of GDA to identify the minimal communication requirements of the relay: we fit an LDA classifier to the fault data and extract the features of most importance to fault classification. 
We use these features to specify the grid state (or topology) changes that the MPMC needs to communicate to the relay to enable adaptive protection.
A stage 2 LDA parametrizes the fault data to statistics that can be used in an optimization formulation to compute relay settings for relay selectivity and coordination.
The setting groups are stored in the memory of each relay IED. 
Depending on the system state, the relay will operate in a certain mode with an associated setting group. 
Post-deployment, the MPMC communicates with the relay to switch its setting group only if the relay needs to adapt in response to a system change.

In the case of a fault, the relay computes its operating time based on local measurements and its relay setting group.
The steps are explained in detail in the following subsections.
We focus the section on the protection against 3-phase faults as the strategy can be easily generalized to the protection against other types of faults. We discuss applicability to different fault types in the Discussions in Section \ref{sec:results}.

\subsection{Offline Steps}

This subsection details the steps that are performed using offline simulation studies prior to deployment.

\textbf{Generate Fault Data.} Monte Carlo simulations {\cite{hammersley2013monte}} are performed to simulate various fault scenarios by leveraging probability distributions over factors that affect measured fault currents and voltages, including grid topology and generator availability and control, and uncertain factors such as fault impedance, fault distance from the relay, renewable energy intermittency, and fault class. 
Power system studies, historical data, or experts' subjective judgments should define these distributions. 

A fault datum for relay $r$, of the form, 
\begin{equation*}
    \{{\mathcal{X}}, \psi, I_F, V_F\}^{(r)}
\end{equation*}
consists of discrete measurements, ${\mathcal{X}}$, that can be infrequently communicated to relay $r$, including switch states (closed/open) and generator modes (in-service/out-of-service); 
the fault current, $I_F$, and voltage, $V_F$, measured by the relay; 
and the fault class, $\psi$. 
In this work, we consider three classes for $\psi$ grouped in the set $\Psi = \{\text{primary}, \text{backup}, \text{other}\}$; but these classes can be extended. 

\textbf{Specify Relay Communication.} In the proposed method, each relay is an agent making decisions based on its local current and voltage measurement and a subset of the measurements ${\mathcal{X}}$ communicated to the relay from the MPMC.
It is desirable to reduce this subset to minimize the reliance (in terms of communication frequency and bandwidth) on communication infrastructure for protection decisions. 
To address this communication requirement, we use the feature importance, as determined by LDA, to specify which measurements need to be communicated to the relay to adapt its settings.
A (stage-1) LDA classifier fits Gaussian distributions to each class of $\psi$ over the space (${\mathcal{X}}$, $I_F$, $V_F$). i.e., 
\begin{equation}
    (I_F, V_F) | \psi \sim \mathcal{N}(\bm{\mu}^{(r)}_\psi, \bm{\Sigma}^{(r)}), \quad \psi \in \Psi 
\end{equation}
after which the subset of ${\mathcal{X}}$ with significant coefficients in $\{ \left( \bm{\Sigma}^{(r)} \right)^{-1} \bm{\mu}^{(r)}_\psi\}^{(\psi \in \Psi)}$ is considered. 
We denote the reduced set $\mathcal{X}^{(r)}_S \subseteq \mathcal{X}$.

Each of the elements in $\mathcal{X}^{(r)}_S$ is an equipment status, which when changed will require the relay to use a different relay setting group.
Hence, $\mathcal{X}^{(r)}_S$ contains information of the minimal distinct modes of operation of the relay.
Grid changes that are not in $\mathcal{X}^{(r)}_S$ would not significantly affect the protection decisions of relay $r$, and hence, are not communicated to the relay.

\textbf{Parametrize Fault Data.} Next, to parametrize the fault data into statistics that we can use for relay optimization and coordination, a (stage-2) LDA fits Gaussian distributions to each class of $\psi$ over $(I_F, V_F)$ for each grid topology identified  in the reduced space $\mathcal{X}^{(r)}_S$. 
For each relay mode $\bm{Z}$, the distributions fitted to the data are:
\begin{align}
    P(I_F, V_F | \psi, \bm{Z}) &= \mathcal{N} \left( I_F, V_F; \bm{\mu}^{(Z)}_\psi, \bm{\Sigma}^{(Z)}_\psi \right), \\
    & \psi \in \Psi, \bm{Z} \in \mathcal{X}^{(r)}_S \nonumber
\end{align}
where 
\begin{equation*}
    \bm{\mu}^{(Z)}_\psi = \begin{bmatrix} 
    {\mu}\{I_F\}^{(Z)}_\psi, {\mu}\{V_F\}^{(Z)}_\psi
    \end{bmatrix}
\end{equation*}
\begin{equation*}
    \bm{\Sigma}^{(Z)}_\psi = \begin{bmatrix} 
    {\sigma}^2\{I_F\}^{(Z)}_\psi & {\sigma}\{I_F,V_F\}^{(Z)}_\psi\\
    {\sigma}\{I_F,V_F\}^{(Z)}_\psi & {\sigma}^2\{V_F\}^{(Z)}_\psi
    \end{bmatrix}
\end{equation*}
denote the mode's fault statistics: ${\mu}\{I_F\}^{(Z)}_\psi$ and ${\mu}\{V_F\}^{(Z)}_\psi$ are the means of the fault current and voltage, respectively, and ${\sigma}^2\{I_F\}^{(Z)}_\psi$ and ${\sigma}^2\{V_F\}^{(Z)}_\psi$ are the variances of the fault current and voltage, respectively, in mode $\bm{Z}$.

\textbf{Compute Relay Settings.} Next, relay settings are optimized, using the above {fault statistics}, to achieve speed and selectivity of protection, and facilitate relay coordination. 
We will describe the parameters of the optimization problem before formulating it.
To simplify relay optimization, we assume $L_i = L_v = 0$ in \eqref{eq:toc} and introduce the variable $\zeta = \text{TMS} \times \lambda_i \lambda_v$ to  derive the following equation for relay operating time: 
\begin{align} \label{eq:toc2}
    t_{op}&(I_M, V_M; \zeta, \eta_i, \eta_v, I_S) =  \\
    & \frac{\zeta}{ 
    \left( \left( \frac{I_M}{I_S} \right)^{\eta_i} - 1 \right)
    \left( \left( \frac{V_S}{V_M} \right)^{\eta_v} - 1 \right)} \nonumber
\end{align}
The objective is to optimize the parameters $\zeta$, $\eta_i$, and $\eta_v$. 

To include information about both the mean and variance of the current and voltage fault data in the optimization problem, we replace $I_M$ and $V_M$ in \eqref{eq:toc2} with 
\begin{equation}
    I_M = \text{CVaR}_{\alpha_i}(I_F) = \mu\{I_F\}^{(Z)}_\psi - \sigma\{I_F\}^{(Z)}_\psi \frac{\phi(\Phi^{-1}(\alpha))}{\alpha}
\end{equation}
\begin{equation}
    V_M = \text{CVaR}_{\alpha_v}(V_F) = \mu\{V_F\}^{(Z)}_\psi - \sigma\{V_F\}^{(Z)}_\psi \frac{\phi(\Phi^{-1}(\alpha))}{\alpha}
\end{equation}
Using the CVaR enables more control over the parametrization of the fault statistics, as we will illustrate in Section \ref{sec:results}.

We use the following notation in the optimization problem formulation: $\mathcal{R}$ is the set of all relays in the system, $\mathcal{B}$ is the set of all buses in the system, $\mathcal{P}_B \subset \mathcal{R}$ is the set of all primary relays to bus $B$, and $\mathcal{S}_p \subset \mathcal{R}$ is the set of all backup relays to relay $p$. $\bm{\zeta}$ is the vector of all settings $\zeta^{(r)}, r \in \mathcal{R}$. $\eta_i, \eta_v$ are the same for all relays in the system for a specified topology.
{
Primary relays should operate following a short delay $\text{D}_{\text{min}}$, and backup relays should operate following an MCT delay. 
}

For all grid topology configurations, we solve the following:
\begin{align} \label{eq:optimization_problem}
    \min_{\bm{\zeta}, \eta} \quad & \sum_{B \in \mathcal{B}} \bigg{(} \sum_{p \in \mathcal{P}_B} \Big{(} t_{op} \big{(} \text{CVaR}^{(p)}_{\alpha_i}, \text{CVaR}^{(p)}_{\alpha_v}; \zeta^{(p)} \big{)} \nonumber\\
    & + \sum_{b \in \mathcal{S}_p} t_{op} \big{(} \text{CVaR}^{(b)}_{\alpha_i}, \text{CVaR}^{(b)}_{\alpha_v}; \zeta^{(b)} \big{)} \Big{)} \bigg{)}
\end{align}
\begin{align}
    & \text{subject to} \nonumber\\
    & \eta_{i,\text{min}} \leq \eta_i \leq \eta_{i,\text{max}}\\
    & \eta_{v,\text{min}} \leq \eta_v \leq \eta_{v,\text{max}}\\
    & \zeta_{\text{min}} \leq \zeta^{(r)} \leq \zeta_{\text{max}}, \quad r \in \mathcal{R}\\
    & t_{op} \big{(} \text{CVaR}_{\alpha_1}; \zeta^{(p)} \big{)} \geq \text{D}_{\text{min}}\\
    & t_{op} \big{(} \text{CVaR}_{\alpha_2}; \zeta^{(b)} \big{)} - t_{op} \big{(} \text{CVaR}_{\alpha_1}; \zeta^{(p)} \big{)} \geq \text{MCT},\\ 
    & \quad \quad \quad \quad B \in \mathcal{B}, p \in \mathcal{R}_B, b \in \mathcal{S}_p \nonumber
\end{align}

Next, relay settings $\{{\zeta}^{(r)}, \eta_i, \eta_v\}$ are stored in-memory on relay $r$ if the grid topology is identified in $\mathcal{X}^{(r)}_S$. 

This optimization problem over $\{\bm{\zeta}, \eta_i, \eta_v\}$ is not convex. However, pre-setting $\eta_i, \eta_v$ yields a linear program over $\bm{\zeta}$. We optimize $\eta_i, \eta_v$ via gradient descent by re-running the linear program for different values of $\eta_i, \eta_v$. While not convex, in practice we have found that starting from very small $\eta_i = \eta_v = 0.01$ yields good optimization results. 

The pickup settings $I_S, V_S$ are preset before optimization. $I_S$ should exceed the maximum full-load line currents, and $V_S$ should be less than bus voltages seen during normal operation.
We compute $I_S$ separately for each relay depending on its line's maximum load, and preset $V_S$ to $0.9$ pu for all relays. 

\subsection{Online Operation}

During operation, each relay will operate in a specific relay mode with a relay setting group, based on the current state of the system. If the system state changes, and the relay's fault response needs to be switched accordingly, the MPMC will communicate with the relay to instruct it to switch to a different relay setting group.
The communication between the MPMC and the relay should be developed to ensure that the MPMC message has been delivered successfully to the relay (e.g., acknowledgments); else, the message should be redelivered.

In the event of a fault, the relay detects the fault by checking whether its local current and voltage drop measurements exceed its pickup settings. Applying (\ref{eq:toc2}), the relay will calculate its operating time. If the fault is not cleared before the relay's operating time is reached, the relay will trip its circuit breaker. If the relay fails to trip, backup relays will have computed a longer operating time and will trip their circuit breakers following an MCT delay. 
After tripping its circuit breaker, the relay will communicate to the MPMC the open status of its circuit breaker.

\section{Simulation Results, Validation and Discussion} \label{sec:results}

\begin{figure}[t]
    \centering
    \vspace{2mm}
    \includegraphics[width=0.8\columnwidth]{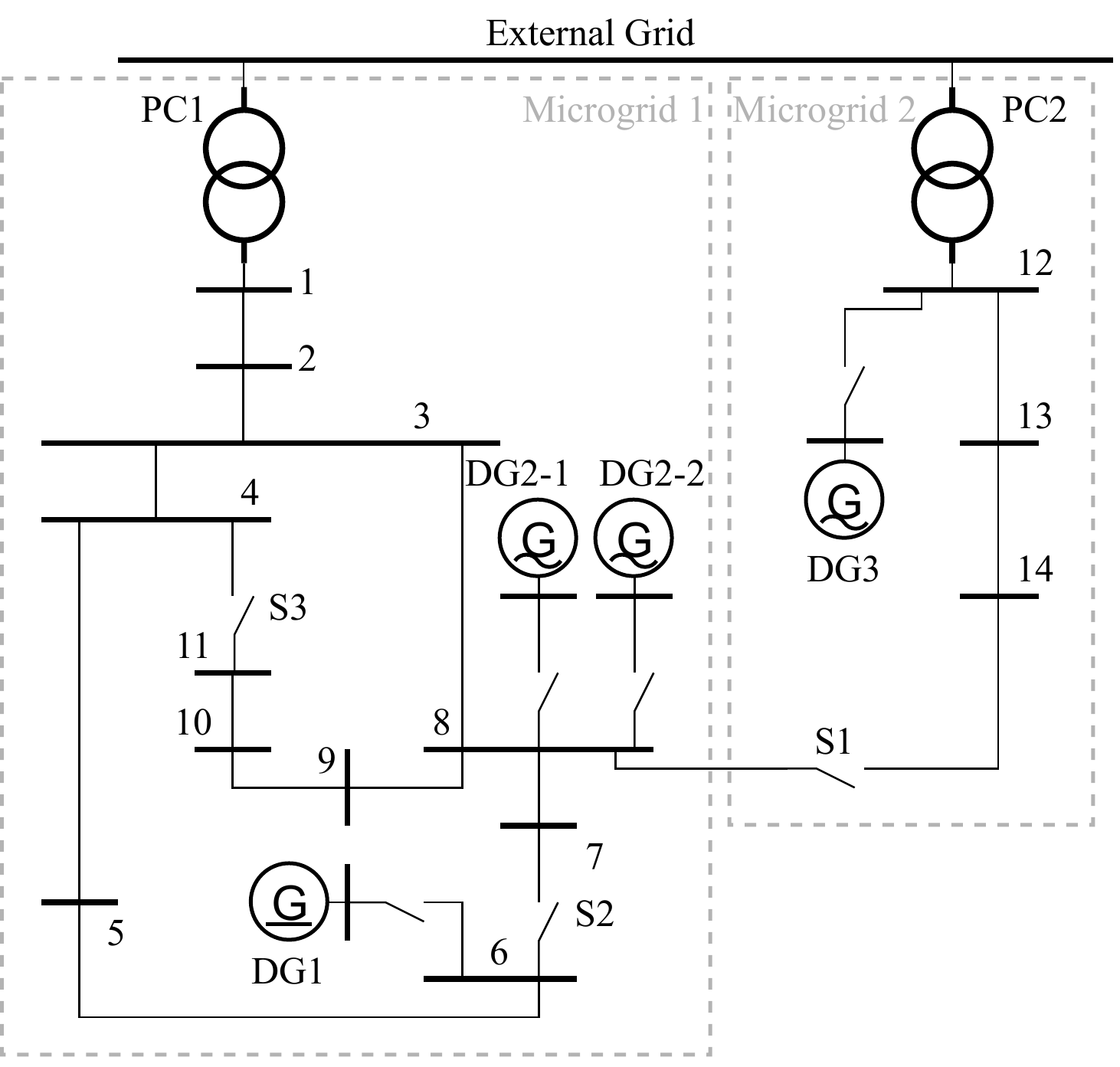}
    \caption{CIGRE medium voltage distribution network \cite{barsali2014benchmark, hooshyar2017microgrid}. It consists of two microgrids (MG). S denotes a switch, PC a point-of-common coupling, and DG a distributed generator. }
    \label{fig:cigre_testbed}
\end{figure}

{
In this section, we will illustrate the proposed method and discuss its benefits. 
}
The benchmark system we use to demonstrate the results is the CIGRE medium voltage distribution network, shown in Fig. \ref{fig:cigre_testbed}. 
The system is composed of two radial microgrids (or feeders), but meshed topology can be obtained by closing switches S1, S2, and S3. 
The system-rated voltage and frequency are 12.47 kV and 60 Hz. 
For the purpose of studying fault current contributions from distributed generation, we add, as in \cite{hooshyar2017microgrid}, three DG units:
\begin{enumerate}
    \item DG1 is a 4 MW rated full-scale VSC-interfaced DG. A 5 MVA, 4.16/12.47 kV, $X=0.07$ pu dYG transformer connects the DG to bus 6. 
    \item DG2 is composed of two 4.3 MVA synchronous generators. A 10 MVA, 4.16/12.47 kV, $X=0.07$ pu dYG transformer connects the DGs to bus 8.
    \item DG3 is a 4.3 MVA synchronous generator. A 5 MVA, 4.16/12.47 kV, $X=0.07$ pu dYG transformer connects the DG to bus 12.
\end{enumerate}

The two microgrids can each operate in islanded mode as DG2 and DG3 can supply entire loads of their respective microgrids, form an islanded mesogrid, or operate in grid-connected mode.
Hence, the benchmark system permits many configurations to help validate the proposed adaptive protection method.
We generate short circuit data using Pandapower \cite{pandapower}.
Pandapower is an open-source software library for power system analysis and optimization in Python, which we use for the ease of integration with Python's data analysis and machine learning libraries.
Pandapower's short circuit calculations are based on IEC60909 \cite{iec60909} and tested against DigSilent.
We also verify our results using a PSCAD time-domain fault simulation.
The probability distributions used to generate Monte Carlo simulations for faults with different grid topology and fault impedance are outlined in Table \ref{table:monte-carlo-vars} in the Appendix. 

The studied cases focus on relay R12-13 and the protection of bus 8 for the insights that can be gained from their analysis. 
The study of R12-13 validates the method for adjusting relay settings in response to variations in fault current resulting from transitioning from an isolated to a grid-connected mode of operation, as well as modifications in network structure from radial to mesh configuration.
The study of Bus 8 faults validates the method for coordinating relays in a mesh topology.
In optimizing the relay settings, we set $\text{D}_{\text{min}} = 3$ cycles, $\text{MCT} = 0.15$ seconds, and $\alpha_i = \alpha_v = 1$.

\subsection{Computing offline relay settings of R12-13}

\begin{figure}
    \centering
    \begin{tabular}{c c}
         \includegraphics[width=0.45\textwidth]{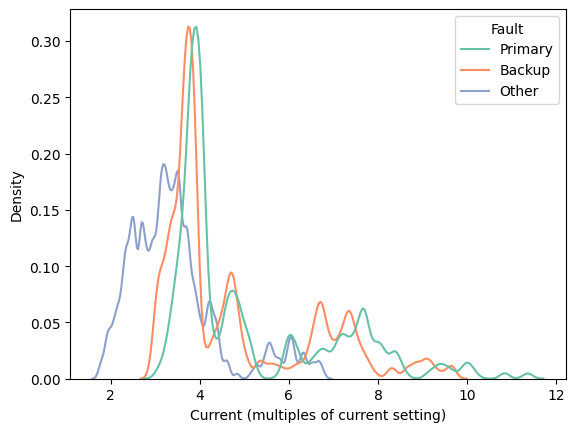}
        &
        \includegraphics[width=0.45\textwidth]{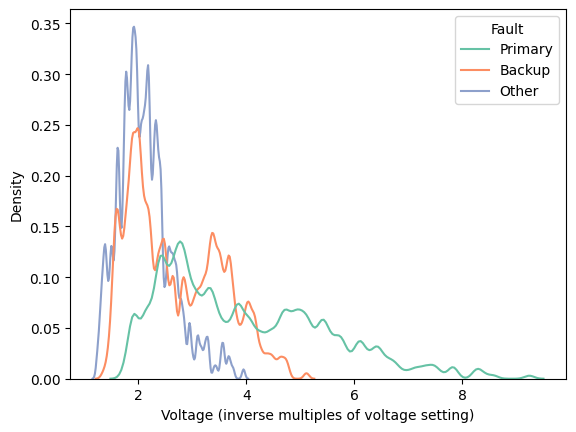}
        \\
        (a) & (b)
    \end{tabular}

    \begin{tabular}{c}
         \includegraphics[width=0.45\textwidth]{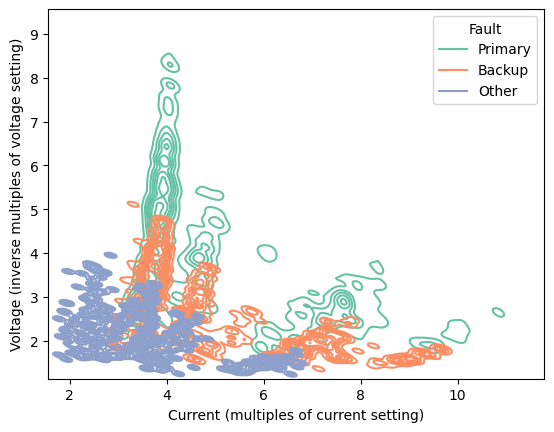}
        \\ (c)
    \end{tabular}
    
    \caption{Kernel density estimation plots of (a) current multiples, (b) voltage inverse multiples, and (c) current multiples and voltage inverse multiples measured by relay R12-13 during 3-phase faults.}
    \label{fig:kdes}
\end{figure}

Fig. \ref{fig:kdes} illustrates kernel density estimations (KDE) of the fault current and voltage data of relay R12-13. The independent variables in Fig. \ref{fig:kdes} (a) and (b) represent the fault currents and voltages measured by R12-13 during the primary, secondary, and other classes of faults. Fig. \ref{fig:kdes} (c) combines the currents and voltages into a single KDE plot.
Note that the current (voltage) is expressed in terms of multiples (inverse multiples) of the current (voltage) pickup setting.
The significant overlap between the different fault classes hinders the ability to classify faults based on the fault current and/or voltage magnitude.
In order to address this issue, the stage-1 LDA separates the relay operation into modes that better distinguish the fault classes based on current and/or voltage measurement to enable a local measurements-based approach for protection.

\begin{figure}
    \centering
    \includegraphics[width=0.4\textwidth, trim={0cm 2cm 17cm 2cm}, clip]{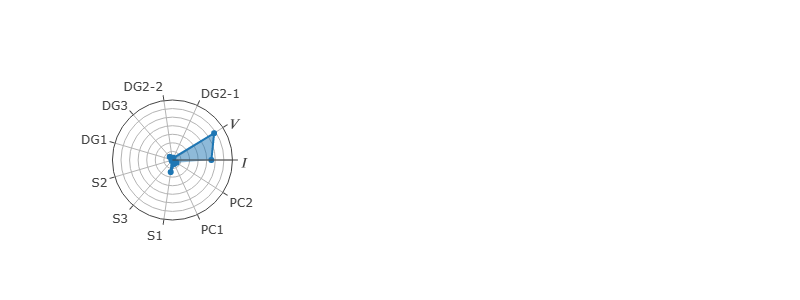}
    \caption{Feature importance coefficients of the GDA classifier trained on R12-13 fault data. The voltage (V) and current (I) are the only two local measurements. The rest are states of switches (S), points-of-common coupling (PC), or distributed generators (DG) that can be communicated to the relay from the MPMC.} 
    \label{fig:star}
\end{figure}

\begin{figure}
    \centering
    \includegraphics[width=0.6\textwidth]{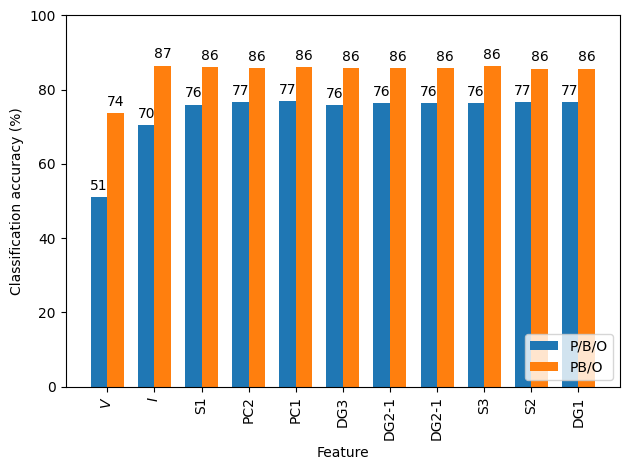}
    \caption{GDA fault classification accuracy. The classes are primary (P), backup (B), and other (O).
    Each bar represents the classification accuracy when the GDA is trained on its feature, in addition to all features to the left of the bar. The right-most bar is the GDA accuracy when trained on all features. The accuracy plateaus after the first 4 features.}
    \label{fig:GDAaccuracy}
\end{figure}

Fig. \ref{fig:star} displays the feature importance coefficients as determined by the stage-1 LDA.
From Fig. \ref{fig:star}, it is observed that the features with the most significant importance coefficients are the voltage and current magnitudes, and the status of the downstream switch S1.
This is further confirmed in Fig. \ref{fig:GDAaccuracy} by evaluating the LDA's classification accuracy as the number of features it has access to is increased (in accordance with the order of feature importance in Fig. \ref{fig:star}).
The blue bars in Fig. \ref{fig:GDAaccuracy} indicate the accuracy when classifying faults into 3 classes (primary, secondary, or other), and the orange bars indicate the accuracy when classifying faults into 2 classes (primary/secondary or other).
The classifier's accuracy (in blue) reaches a plateau of $77\%$ after being provided with a small subset of the measurements comprising the voltage and current magnitudes, and the states of S1 and PC2.
As such, we compartmentalize the R12-13's operation into $2^2 = 4$ modes of operation corresponding to permutations of each of S1 and PC2 being open or closed.
Hence, the MPMC needs only to communicate the status of S1 and PC2 to relay R12-13.

\begin{figure}
    \centering
    \begin{tabular}{ >{\centering\arraybackslash} m{0.02cm} >{\centering\arraybackslash} m{6cm} >{\centering\arraybackslash} m{6cm}}
        {\centering (a)} & \includegraphics[width=0.37\textwidth]{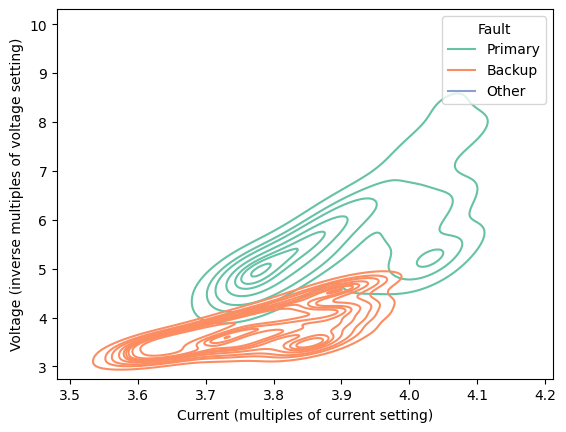}
        & 
        \includegraphics[width=0.37\textwidth]{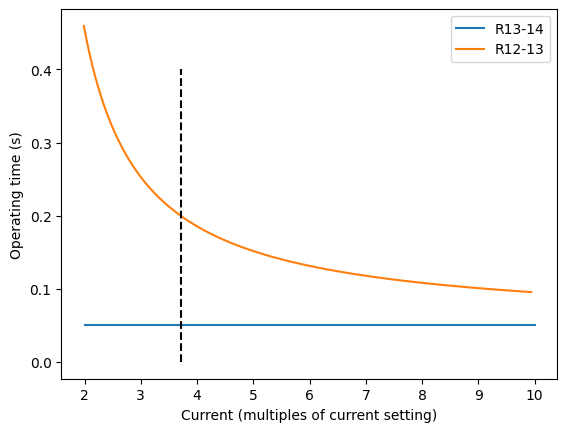}
        \\
        
        (b) & \includegraphics[width=0.37\textwidth]{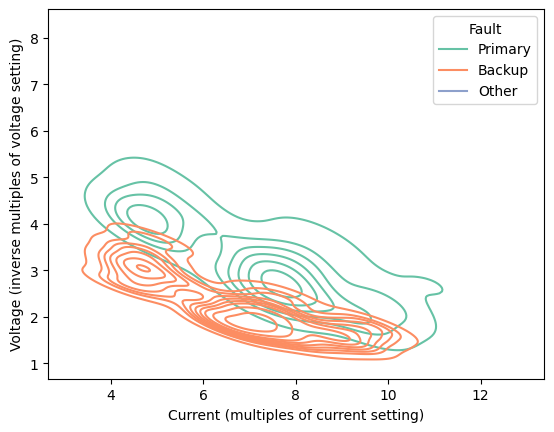}
        &
        \includegraphics[width=0.37\textwidth]{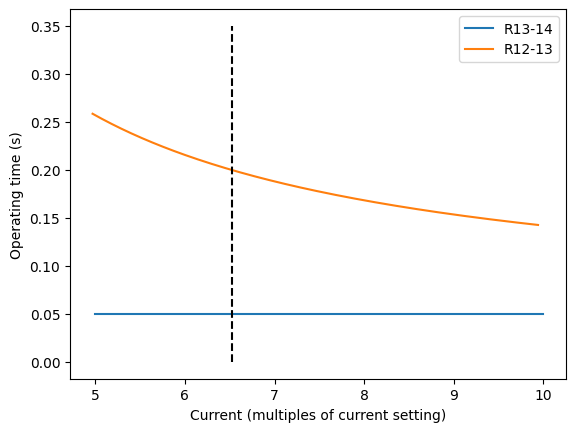}
        \\
        
        (c) & \includegraphics[width=0.37\textwidth]{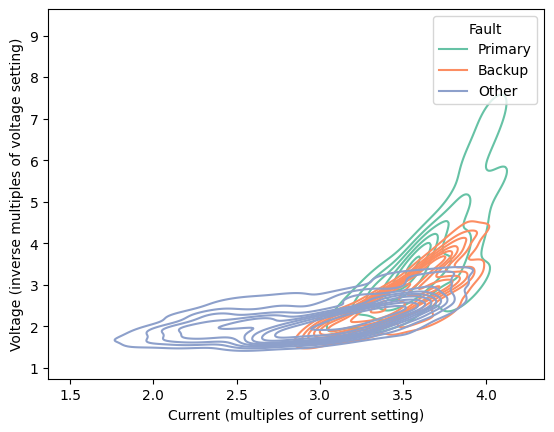}
        &
        \includegraphics[width=0.37\textwidth]{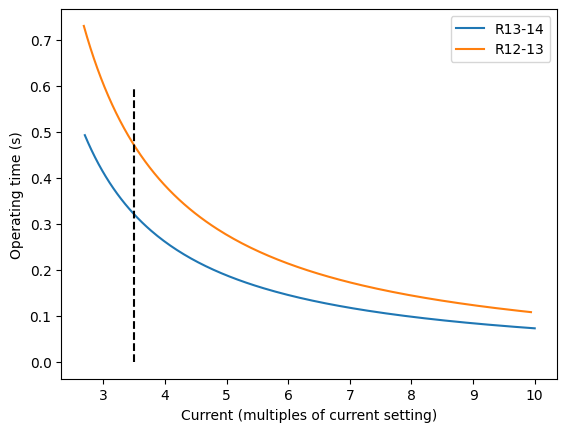}
        \\
        (d) & \includegraphics[width=0.37\textwidth]{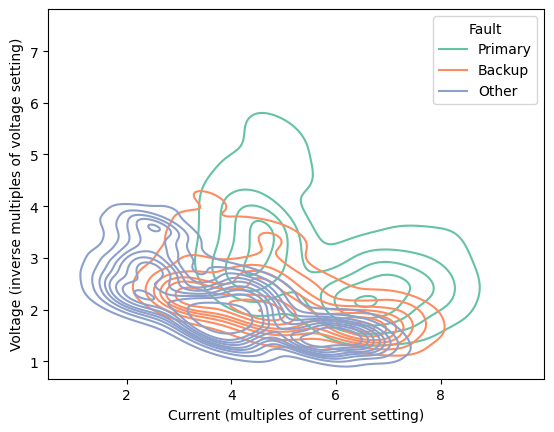}
        &
        \includegraphics[width=0.37\textwidth]{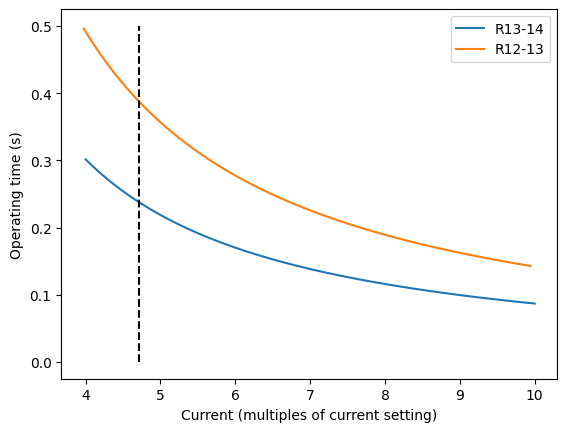}        
    \end{tabular}
    \caption{(left) Kernel density estimation plots of current multiples and voltage inverse multiples measured by relay R12-13 and (right) relay operating curves against current multiples for each of its four modes. Modes (a): S1, PC2 open, (b): S1 open, PC2 closed, (c): S1 closed, PC2 open, (d): S1, PC2 closed.}
    \label{fig:r1213States}
\end{figure}

In Fig. \ref{fig:r1213States} (a) to (d), we divide R12-13's operation into these 4 modes and plot the KDEs of the fault current and voltage data in each mode.
Fig. \ref{fig:r1213States} (a) and (b) correspond to the modes with S1 open; hence, R12-13 only measures primary and secondary faults.
Fig. \ref{fig:r1213States} (b) and (d) correspond to the modes with PC2 closed; hence, R12-13 experiences higher fault current magnitudes due to the increased contribution from upstream sources, including the main grid.

It can be observed that by compartmentalizing the relay operation into modes, the 
overlap between the different fault classes is significantly reduced, facilitating the distinction of fault classes based on voltage and current magnitude.
The fault current (voltage) magnitude decreases (increases) as the distance from the fault increases.
This enables the use of inverse-time protection to coordinate relays.
The objective of the stage-2 LDA is to parametrize the fault data so that their distribution statistics can be utilized for relay coordination.

\begin{figure}
    \centering
    \includegraphics[width=0.6\textwidth]{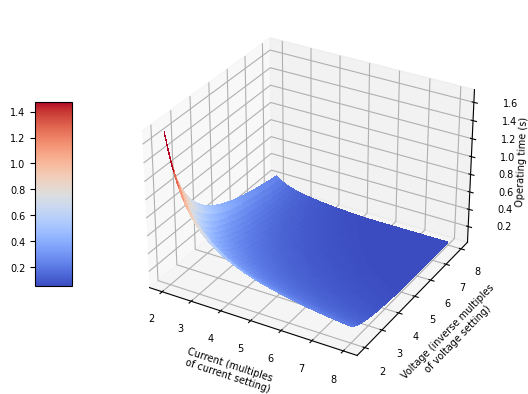}
    \caption{Relay operating curve of relay R12-13 against current multiples and voltage inverse multiples for mode (c): S1 closed, PC2 open.}
    \label{fig:topVI}
\end{figure}

To the right of each KDE plot in Fig. \ref{fig:r1213States}, we plot the corresponding inverse-time operation curves against the fault current (measured as multiples of pickup current setting). 
While the relay operation time is a function of both the fault current and voltage, we only plot the operation curves against the fault current for the purposes of demonstrating relay coordination, as it is hard to observe the coordination in a 3-dimensional plot.
We illustrate the full 3-dimensional inverse-time operation curve of R12-13 against both fault current and voltage (measured as inverse multiples of pickup voltage setting) for mode (c) in Fig. \ref{fig:topVI}.

To demonstrate relay coordination:
\begin{itemize}
    \item We plot vertical dashed lines in Fig. \ref{fig:r1213States} to illustrate the CVaR of primary fault currents measured by R13-14 (blue), i.e., $\text{CVaR}_{\alpha_i}(I_F)^{13-14}$. 
    \item Since the current multiple measurements by the backup relay R12-13 might be different than that measured by R13-14 due to different $I_S$ settings or line branching, we shift the operation curve of R12-13 so that the vertical line also matches how R12-13 measures the fault current.
    \item The voltage value of each of the operation curves of R13-14 and R12-13 corresponds to the voltage values each measures during the fault characterized by the vertical line.
\end{itemize}
The intersection of the relay curve with the dashed line marks the relay operation time; hence, the higher curve trips later.
Relay coordination can be easily evaluated by looking at the relay operation curves. R12-13 serves as a backup for R13-14, which can be observed from R12-13's curve being higher than R13-14's curve in all modes. 

Fig. \ref{fig:r1213States} illustrates how the relays adjust to changes in the grid operation, including going from connected to islanded modes and switching between radial and meshed topologies. 
The 4 relay group settings for R12-13 and its downstream relay R13-14 are included in Table \ref{table:m2settings} in the Appendix. These settings groups must be stored in their memories to enable them to adapt to changing conditions.

\textit{Remarks:} In modes (a) and (b) when S1 is open, the optimization results in a TMS of 0 for R13-14, which means it is sufficient to operate R13-14 as an instantaneous over-current relay. The increase in operation time for the relays when S1 is closed allows for coordination with relays further downstream.

In Fig. \ref{fig:r1213States} (c), an important insight is the distinctive `crescent' shape observed in the KDE plot. This shape indicates that voltage drops increase more significantly than fault current increases for nearer faults. 
A factor contributing to this is the limited fault current of distributed generation.
As a result, incorporating voltage measurements enhances the ability to distinguish fault classes in this mode. This finding supports our decision to include voltage drops in the inverse-time relay operation.

\begin{figure}
    \centering
    \includegraphics[width=0.55\textwidth]{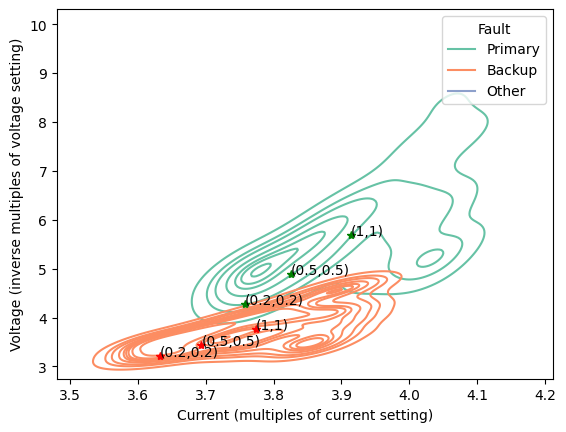}
    \caption{Illustrating the benefits of CVaR in optimization. Each point represents $(\text{CVaR}_{\alpha_i}(I_F)^{12-13}, \text{CVaR}_{\alpha_v}(V_F)^{12-13})$. The values of $(\alpha_i, \alpha_v)$ are annotated.}
    \label{fig:cvar}
\end{figure}

Lastly, we discuss how the use of CVaR is beneficial in the relay optimization process.
Fig. \ref{fig:cvar} shows how different value pairs of CVaR for current and voltage can be used to better control how the data is parametrized before it is used in the optimization. Lowering the CVaR values allows for greater emphasis on the lower tails of the fault current and voltage distributions. This is useful because inverse-time protection is faster at clearing faults with high current and low voltage magnitudes (top-right corner of the figure). Lowering (raising) the CVaR values shifts the focus of the optimization and allows for faster (slower) clearing of faults.

\subsection{Relay coordination at Bus 8 with topology change}

\begin{figure}
    \centering
    \begin{tabular}{c c}
        \includegraphics[width=0.45\textwidth]{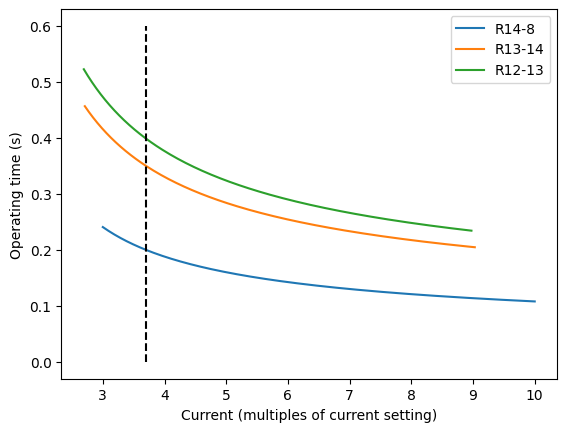}
        &
        \includegraphics[width=0.45\textwidth]{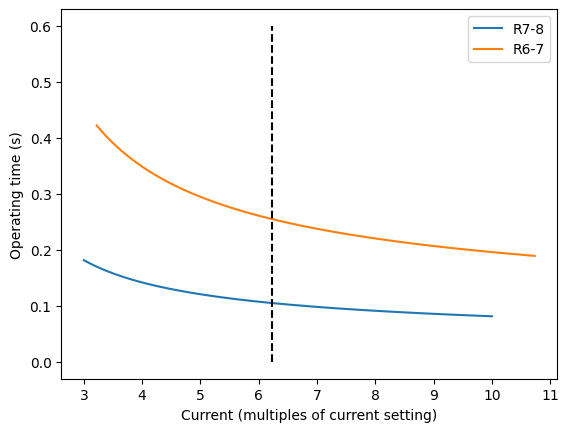}
        \\
        (a) & (b)\\
        \includegraphics[width=0.45\textwidth]{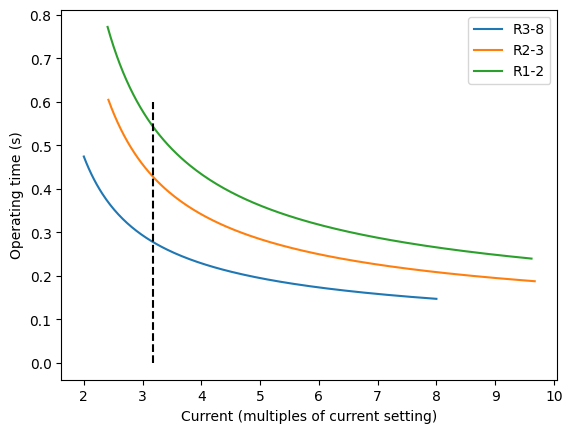}
        &
        \includegraphics[width=0.45\textwidth]{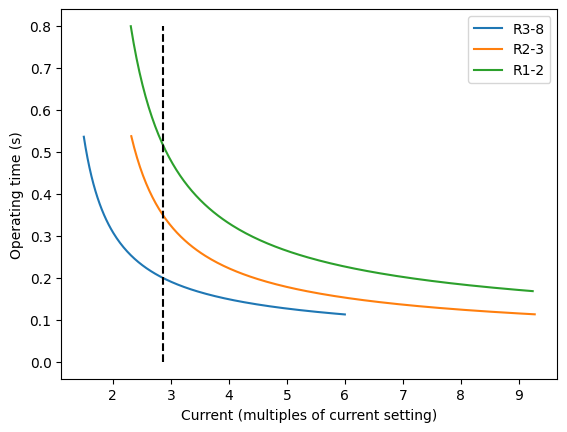}
        \\
        (c) & (d)
    \end{tabular}
    \caption{
    Relay operation curves against current multiples for relays (a) R14-8, R13-14, and R12-13, (b) R7-8 and R6-7, and (c) R3-8, R2-3, and R1-2 for a fault at bus 8 when S3 is open and S1, S2, PC1 and PC2 are closed.
    (d) Relay operation curves against current multiples for relays R3-8, R2-3, and R1-2 for a fault at bus 8 when S1, S2, and S3 are open and PC1 is closed.}
    \label{fig:r38States}
\end{figure}

\begin{figure}
    \vspace{2mm}
    \centering
    \includegraphics[width=0.7\textwidth]{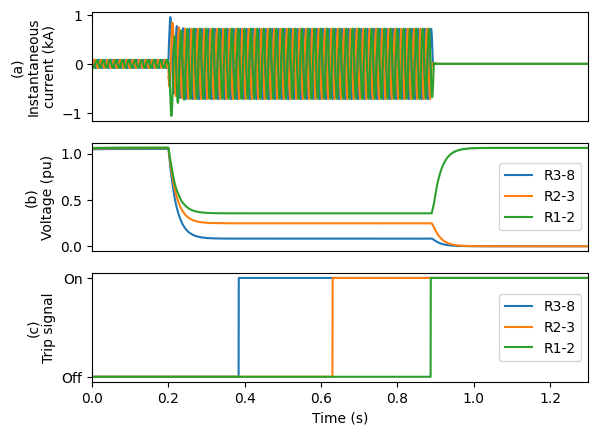}
    \caption{PSCAD simulation of the (a) instantaneous current measured by R1-2, (b) voltages, and (c) tripping signals of R3-8, R2-3, and R1-2 for a $1 \Omega$ fault at bus 8 with PC1 closed, and switches S1, S2, and S3 open.}
    \label{fig:pscad}
\end{figure} 

To validate the method for coordinating relays in a mesh topology, we consider faults at bus 8. When S1 and S2 are closed, a fault at bus 8 is isolated by three primary relays: R14-8, R7-8, and R3-8. 
Each primary relay has one or more backup.

In Fig. \ref{fig:r38States}, we plot the relay operation curves of relays (a) R14-8, (b) R7-8, and (c) R3-8 and their backup relays when PC1, PC2, S1 and S2 are closed. 
The dashed vertical lines show the CVaR of fault currents measured by the primary relays for faults at bus 8. 
The figure demonstrates the successful isolation of the fault from all generation sources and relay coordination in case of relay failure.

If S1 and S2 are open, the topology of the microgrids becomes radial with R3-8 offering primary protection to bus 8 and R2-3 and R1-2 its backups. 
The fault current is supplied by the main grid. 
The relay operation curves of these relays in this case are shown in Fig. \ref{fig:r38States}(d).
PSCAD simulations validate the coordination results of this case study. In Fig. \ref{fig:pscad}, a $1 \Omega$ fault occurs at bus 8 at the 0.2 seconds mark. We suppress the operation of the primary (R3-8) and its backup (R2-3) relays, and let R1-2 isolate the fault. The operation times of the relays closely match the coordination curves in Fig. \ref{fig:r38States}(d).

The relay settings for these two case studies are listed in Table \ref{table:m1settings} in the Appendix.

\subsection{Discussion}

{
The proposed method has the following merits:}

\begin{figure}
    \centering
    \includegraphics[width=0.6\textwidth]{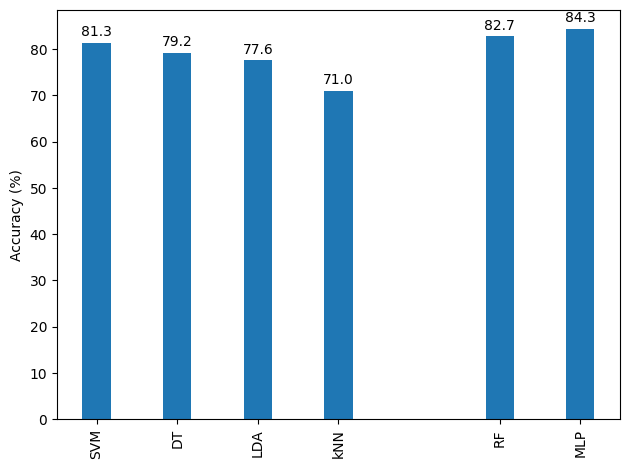}
    \caption{Comparing LDA classification accuracy with other algorithms.}
    \label{fig:compareAccuracy}
\end{figure}

\subsubsection{Evaluating LDA accuracy}

The significant overlap of different fault classes observed in Fig. \ref{fig:kdes} shows that it is infeasible to classify a fault based only on local current and voltage measurements.
In Fig. \ref{fig:compareAccuracy}, we compare the classification accuracy of multiple machine learning algorithms in classifying the faults measured by R12-13 when limiting the training data to discrete grid states and local measurements -- to avoid the reliance on real-time communication.
On the left side of Fig. \ref{fig:compareAccuracy}, we include algorithms that yield interpretable results, including SVM, DT, LDA, and kNN. 
On the right, we include algorithms that generally have lower learning bias and higher accuracy, but do not yield interpretable results, including random forest (RF) and multi-layer perceptrons (MLP). 
Hyperparameter optimization for these algorithms is listed in Table \ref{table:hyperparameters} in the Appendix.
We observe that MLPs achieve the highest classification accuracy, as expected;
but, highlighting the difficulty of the classification task, the maximum accuracy is only $84\%$. 

Note that these algorithms have been shown to yield higher performance in centralized real-time communication-based protection strategies in the literature, when not constrained to local measurements and discrete system states. But this exposes the protection to reliability threats related to the communication infrastructure.

While LDAs accuracy is lower, the interpretability of the LDA results enabled us to augment it with an optimization-based approach that facilitates relay coordination, preventing miscoordination between relays and their backups.
The interpretability of LDA enabled easy visualization of the fault data statistics, protection decisions, and relay coordination.
Hence, the results of the method can be easily interpreted and validated by system operators and protection engineers, facilitating its use for adaptive protection and reducing associated implementation risks and reliability threats.

Despite its lower accuracy, we prefer LDA over DT due to its stability. 
Further, we prefer LDA over SVM since LDA enables fitting Gaussian distributions to fault classes, which we use to reduce problem dimensionality and parametrize the fault data for optimization. 
As observed by the results, Gaussian  distributions fit well to the fault data. 

\subsubsection{Communication Network Delays and Cyberattacks}

The proposed protection strategy requires communication of the grid topology to the relays. Generally, this reliance on communication infrastructure can expose the protection relay operation to potential risks, such as network delays or cyberattacks. To address this issue, our method minimizes communication to the relays by communicating only a subset of the infrequent major changes to the grid topology. The necessary settings to operate the relays in real-time are stored locally in memory on the relays. 
This provides an advantage over centralized protection strategies that require a constantly active communication channel with the relay. 
Due to the infrequent communication, temporary network delays will not significantly impact the protection. 
Further, the infrequency permits the use of encryption algorithms to authenticate and protect the communication to relays. This approach ensures that the proposed protection strategy is more secure against potential cyberattacks and/or network delays.

\subsubsection{Achieving Protection Relay Characteristics}

The strategy extends the operation of inverse-time protection relays by adapting protection decisions to changing grid conditions and considering voltages for an improved distinction of faults in systems with limited fault currents. 
Below, we evaluate the proposed strategy against the characteristics of effective protection strategies.

\begin{itemize}
    \item Reliability: 
    The proposed protection strategy detects and clears faults in a timely and accurate manner. 
    The optimization problem minimizes the operation time of relays to expedite isolating faults.
    
    To prevent false alarms or misoperations, the relays in the proposed strategy, 
    similar to existing inverse-time over-current relays, 
    do not trigger unless the fault current exceeds a preset pick-up current setting. 

    Uncontrolled variables, including fault impedance, inception angles, pre-fault power system conditions, and relay measurement errors, can be modeled into the Monte-Carlo simulations.
    Further considerations, including in-rush currents due to switching of loads, distributed generation, filter capacitors, or transformers and transformer saturation, can be addressed as is currently with over-current relays or included as constraints in the optimization \eqref{eq:optimization_problem}.

    \item Selectivity: 
    As illustrated in the KDE plots, the proposed protection strategy compartmentalizes relay operation into modes, in which it is more feasible for the relay to discriminate between different types of faults based on local measurements. 

    \item Coordination:
    As demonstrated by relay curves in multiple case studies in the paper, the protection strategy is able to coordinate relays to isolate minimal sections of the system during faults.   

    Non-adjustable devices, such as fuses, should be coordinated with the relays, or further constraints can be considered in the optimization \eqref{eq:optimization_problem} to coordinate with any devices. 

    \item Adaptability: 
    Cases have demonstrated the ability of the relays to adapt to transitions from isolated to a grid-connected mode of operation with higher short-circuit capacities, as well as modifications to network topology from radial to mesh configuration.
    
    \item Simplicity: 
    Modern IEDs used for protection can readily process the proposed protection strategy.
    The strategy minimizes the reliance on the communication infrastructure and memory use of the IEDs.

    In addition, the method yields interpretable decisions that can be easily validated by protection engineers.

    \item Communication: 
    The strategy relies on a communication infrastructure between the MPMC and relays. 
    Commands can be established from the MPMC to switch the relay setting group by means of the IEC61850 protocol. 
    
    The method minimizes the communication bandwidth and frequency, only relying on infrequent updates due to changes in a grid topology.

\end{itemize}

\subsubsection{Applicability to Different Protection Relays and Fault Types}

The proposed method specifies the communication requirements and optimizes relay settings. 
While we demonstrated our method on inverse-time protection relays, the method can be extended to other protection devices, such as distance relays, to adapt their operation to changing grid conditions. 
We emphasize that the main contribution of the paper is the data-driven component: applying GDA to the adaptive protection of power systems.

We demonstrated the strategy for 3-phase faults.  
The proposed strategy is applicable to other fault types.
In the same way that relay settings were computed for 3 phase faults in the paper, data can be generated and relay settings can be computed for other fault types.
The relay settings will be stored on the relay IED memory. 
During a fault, the relay will use the current and voltage measurement to distinguish the fault type. 
The relay settings for the fault type and relay mode will be used to compute an operation time and isolate the faulted phases.

\section{Conclusion} \label{sec:conclusion}

In recent years, there has been a growing interest in utilizing machine learning for adaptive protection in power systems, leveraging the increasing amount of power system data. However, safety-critical applications such as power system protection require methods that are transparent, interpretable, and easily validated. Deep learning methods, although highly accurate, are difficult to validate -- presenting a significant risk to their utilization for power system protection.

To address this issue, we propose a novel adaptive protection strategy for inverse-time relays based on probabilistic machine learning, specifically Gaussian discriminant analysis. We apply the strategy to distill power system data to inverse-time relay settings, generating operation curves that can be used to verify and validate protection decisions.
We demonstrate the use of our strategy for fault detection and localization, relay coordination and optimization, and protection management by specifying minimal communication requirements for adapting relays. 
To demonstrate the effectiveness of our method, we apply it to the protection of the CIGRE medium voltage grid, as it switches between islanded and grid-connected modes of operation, as well as radial and mesh topology in the presence of distributed generation.

\appendix

\begin{table}[!htb]
\centering
\caption{Monte Carlo Simulation Variations}
\begin{tabular}[t]{lc} 
\toprule
\textbf{Random variable} & \textbf{Values}\\
\midrule
Fault resistance & $\text{Unif}(0, 3) \Omega$ \\
Fault reactance & $\text{Unif}(0, 3) \Omega$ \\
Main grid short-circuit capacity & $20-22$ MVA\\
DG1 short-circuit to nominal current ratio & $1.1-1.3$\\
Switch status & On/Off\\
PC states & On/Off\\
Generator in-service & True/False\\
\bottomrule
\label{table:monte-carlo-vars}
\end{tabular}
\end{table}

\begin{table}
\centering
\caption{Relay Settings for relays R12-13 and R13-14}
\begin{tabular}[t]{lcccc} 
\toprule
\textbf{Mode} & \textbf{$\eta_i$} & \textbf{$\eta_v$} & \textbf{$\zeta^{13-14}$} & \textbf{$\zeta^{12-13}$}\\
\midrule
S1 Open, PC2 Open & 1 & 1 & 0 & 0.81\\
\midrule
S1 Open, PC2 Closed & 1 & 1 & 0 & 0.62\\
\midrule
S1 Closed, PC2 Open & 1.24 & 0.9 & 2.03 & 2.52\\
\midrule
S1 Closed, PC2 Closed & 1.2 & 1.01 & 3.17 & 2.27\\
\bottomrule
\label{table:m2settings}
\end{tabular}
\end{table}

\begin{table}[t!]
\centering
\caption{Settings of relays protecting bus 8}
\begin{tabular}[t]{lccccc} 
\toprule
\textbf{Mode} & \textbf{$\eta_i$} & \textbf{$\eta_v$} & \textbf{Primary} & \textbf{Backup}\\
\midrule
Open: S1, S2, S3 & 0.1 & 2.89 & $\zeta^{3-8} = 5.07$ & $\zeta^{2-3} = 0.66$\\
Closed: PC1, PC2\\
\midrule
Closed: S1, S2 & 0.1 & 4.75 & $\zeta^{3-8} = 7.03$ & $\zeta^{2-3} = 2.26$\\
PC1, PC2 & & & $\zeta^{14-8} = 3.25$ & $\zeta^{13-14} = 1.85$\\
Open: S3 & & & $\zeta^{7-8} = 6.75$ & $\zeta^{6-7} = 14.91$\\
\bottomrule
\label{table:m1settings}
\end{tabular}
\end{table}

\begin{table}
\centering
\caption{Hyperparameter optimization}
\begin{tabular}[t]{lccc} 
\toprule
\textbf{Algorithm} & \textbf{Hyperparameters} & \textbf{Range} & \textbf{Optimal}\\
\midrule
RF & Number of DTs & 5-40 & 30\\
kNN & Number of neighbors & 2-10 & 8\\
MLP & Hidden layers & 5-80 &  40\\
\bottomrule
\label{table:hyperparameters}
\end{tabular}
\end{table}

\bibliography{references}

\end{document}